\documentclass[12pt]{iopart}
\usepackage[dvips]{graphicx}
\begin{document}

\title{Giant magneto-caloric effect near room temperature in Ni-Mn-Sn-Ga alloys}
\author{S. Chatterjee, S. Giri, S. Majumdar$^*$}
\address{ Department of Solid State Physics, Indian Association for the Cultivation of Science, 2A \& B Raja S. C. Mullick Road, Jadavpur Kolkata 700 032, India}
\ead{$^*$sspsm2@iacs.res.in}

\author{S. K. De} 
\address{Department of Materials Science, Indian Association for the Cultivation of Science, 2A \& B Raja S. C. Mullick Road, Kolkata 700 032, INDIA}

\begin{abstract}

We report the observation of giant magneto-caloric effect (MCE) in ferromagnetic shape memory alloys (FSMAs) of nominal compositions Ni$_2$Mn$_{1.36}$Sn$_{0.64-x}$Ga$_{x}$ ($x$ =0.24, 0.28 and 0.32 ). Irrespective of the Ga doping, all the samples undergo long range ferromagnetic ordering below around 330 K. However, the martensitic transition temperature ($T_{MS}$) of the samples varies strongly with Ga concentration. Clear signature of field induced transition around $T_{MS}$ is visible for all the samples. The observed MCE ($\Delta S$ = 13.6 J/kg K for field changing from 0 to 50 kOe) is found to be highest for $x$ = 0.28 with peak near 249 K, while $x$ = 0.32 shows $\Delta S$ = 12.8 J/kg K at 274 K. This series of Ga-doped alloys are found to be interesting materials with high value of $\Delta S$ over a varied range of temperature.
\end{abstract}




\maketitle


\section{Introduction}
The phenomenon of magneto-caloric effect (MCE) has recently attracted considerable attention due to their possible application in magnetic refrigeration~\cite{kag}. Among others, Gd$_5$Ge$_4$-derived compounds show large MCE around the first order mageto-structural transition~\cite{vkp}. Recently, Ni-Mn-Z (Z = Sn, In, Sb) Heusler type ferromagnetic shape memory alloys (FSMAs) have been found to show various magneto-functional properties, which include magnetostriction, magnetoresistance, MCE, exchange bias etc~\cite{sutou, kainuma,vishnu, imce, yu, eb}. The interesting point regarding the MCE of Ni-Mn-Z alloys is that the observed change in entropy under magnetic field ($H$) is high and positive near the martensitic type structural transition. The effect is related to the magnetic field induced structural transition ~\cite{koyama}, which gives rise to large change in entropy~\cite{planes, krenke}.

\par
In order to look for suitable magneto-caloric materials among the FSMAs, various doping studies have been performed in recent times~\cite{liu, sk, rk, pathak}. Here we report the large MCE observed in a solid solutions of Ga and Sn based Heusler alloys. It is known that Ni$_2$Mn$_{1.36}$Sn$_{0.64}$ alloy shows ferromagnetic shape memory effect with martensitic transition (MT) temperature close to 110 K~\cite{sc1}. In the present study, we observe that the gradual replacement of Sn by Ga pushes the martensitic transition temperature ($T_{MS}$) to higher values keeping the ferromagnetic Curie point ($T_C$) almost constant.

\section{Experimental Details}
The polycrystalline samples of nominal compositions Ni$_2$Mn$_{1.36}$Sn$_{0.64-x}$Ga$_{x}$ ($x$ =0.24, 0.28 and 0.32 ) used in the present work were prepared by melting the constituent metals and subsequent annealing. The samples were characterized by powder x-ray diffraction (XRD) and energy dispersive x-ray spectroscopy. At room temperature, the samples are found to be single phase with cubic Heusler structure (L2$_1$). No traces of unreacted Ga peaks were present in the pattern for any of the samples. The cubic lattice parameter ($a_{cubic}$) of the samples is found to decrease linearly with $x$ obeying the Vegard's law for alloy formation (see inset of fig. 1 and table 1). Magnetization ($M$) of the samples were measured using quantum design SQUID magnetometer (MPMS 6, Evercool model) as well as vibrating sample magnetometer from Cryogenic Ltd., UK.

\begin{figure}%
\centering
\includegraphics[width=8.5 cm]{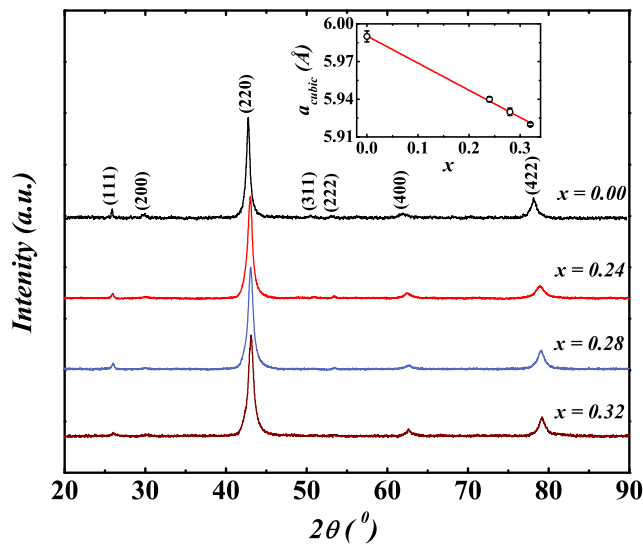}%
\caption{Powder x-ray diffraction pattern at room temperature for the Ni$_2$Mn$_{1.36}$Sn$_{0.64-x}$Ga$_{x}$ alloys using Cu K$_{\alpha}$ radiation. The inset shows the variation of cubic lattice parameter with Ga concentration.}%
\end{figure}

\section{Results}
Fig.2 shows the $M$ versus temperature ($T$) data of the samples measured in different protocols, namely zero-field-cooled-heating (ZFCH), field-cooling (FC) and field-cooled-heating (FCH) under $H$ = 100 Oe. The thermal hysteresis between FC and FCH indicates the first order MT in the sample. $T_{MS}$ of the samples (which is defined as the onset of the drop in $M$ in the FC data with decreasing $T$) is found to increase gradually with increasing $x$. This has been indicated in table 1. The sharp drop in $M$ near the MT is identical to the similar in other Ni-Mn-Z alloys. This indicates that like the undoped samples, Ni$_2$Mn$_{1.36}$Sn$_{0.64-x}$Ga$_{x}$ alloys show lower value of $M$ in the martensitic phase. The ZFC data for all the samples branch out from the MT point, indicating the development of thermo-magnetic irreversibility across the the transition. All the samples are ferromagnetic (FM) below room temperature and $T_C$'s of the samples are found to be close to 330 K (as apparent from the ac susceptibility and resistivity data, not shown here). A further step-like anomaly is observed well below the MT in the ZFC data (indicated by $T_B$), which is also observed in various other Ni-Mn-Z alloys.

\begin{figure}%
\centering
\includegraphics[width=6.5 cm]{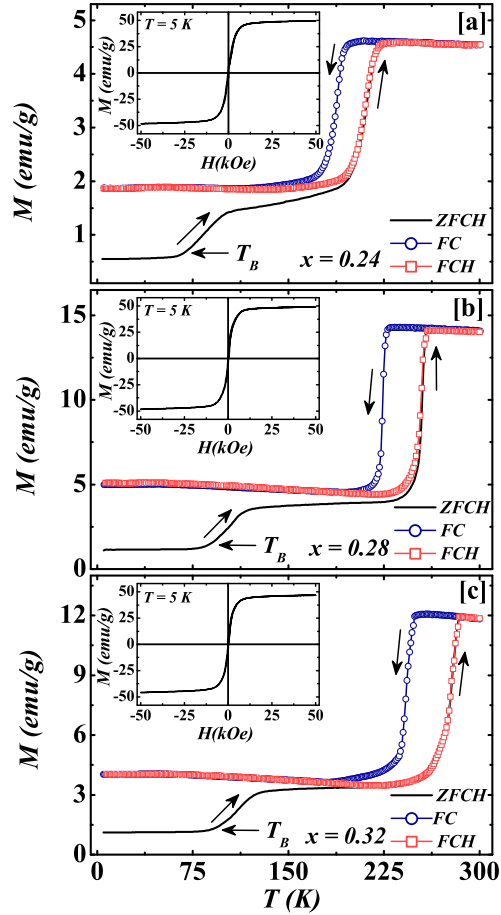}%
\caption{Magnetization as a function of temperature for $x$ = 0.24, 0.28, and 0.32 samples measured in zero-field-cooled heating (ZFCH), field-cooled-heating (FCH) and field-cooling (FC) conditions at an applied field of 100 Oe. The insets show the isothermal magnetization versus field data at 5 K.}%
\end{figure}

\par
We measured isothermal $M$ as a function of $H$ at various $T$. The 5 K isotherm for the samples are shown in the insets of figure 2. Under an applied $H$, all the samples show typical soft ferromagnetic behaviour (coercive field $\sim$ 100 Oe), with sharp rise at low $H$ and saturation at higher fields. The 5 K saturation moment ($M_{sat}$) is found to decrease with increasing Ga concentration (see table 1). In fig. 3, we have shown several isotherms for the representative sample $x$ = 0.28, which is archetype of other compositions. The sample has $T_{MS}$ = 231 K. Outside the region of MT (fig.3 (a) and (c)), the isotherms show tendency toward saturation at 50 kOe of field. However, in the region of the MT (fig. 3 (b)), $M-H$ curves are much complex, with clear signature of field induced transition. At $T$ = 249 K, the metamagnetism is particularly clear, where $M$ starts to rise beyond 30 kOe of field. In comparison with other Ni-Mn-Sn~\cite{sutou} samples, such metamagnetism is likely to be the field-induced reverse transition, which drives the martensite fraction into austenite in the highly metastable region of MT. Similar $M-H$ behaviour is present in other Ga doped samples, as well as in the undoped $x$ = 0 sample (not shown here).

\par
Considering the presence of large metamagnetic transition, we have calculated the MCE of the samples around MT. MCE is often expressed as the change in entropy ($\Delta S$) by the application of $H$, which can be obtained from magnetization data using Maxwell's thermodynamical relation~\cite{kag}: ${\Delta}S(0{\rightarrow}H)=\int^H_0\left(\frac{{\partial}M}{{\partial}T}\right)_HdH$. In order to calculate $\Delta S$, $M$ versus $T$ data at different $H$ were obtained by convoluting $M(H)$ data recorded at different constant $T$ in the thermally demagnetized state. These $M$ versus $T$ data at different $H$ have been depicted in fig. 3 (d) for $x$ = 0.28 sample. The data show the signature of MT, where $M$ changes sharply with $T$. Notably, MT shifts to lower $T$ with increasing $H$.

\begin{figure}%
\centering
\includegraphics[width=12 cm]{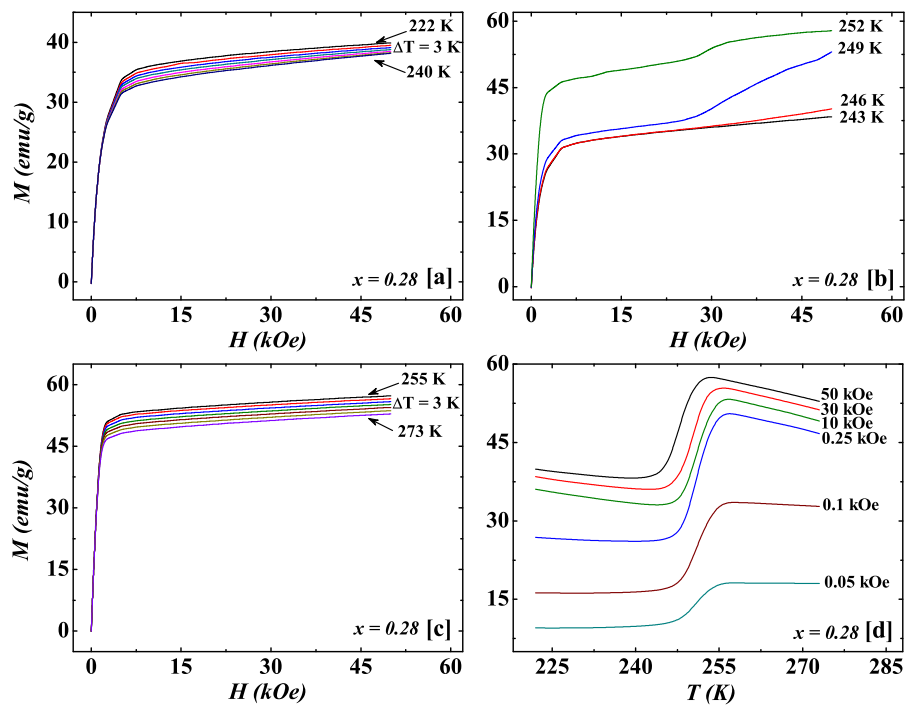}%
\caption{(a), (b) and (c) show isothermal magnetization as a function of field at different temperatures for $x$ = 0.28 sample measured in thermally demagnetized state. (d) shows the convoluted magnetization versus temperature data at different fields obtained from the $M-H$ isotherms.}%
\end{figure}

\par
We have calculated $\Delta S$ for $H$ changing from 0 to different final values (10, 20, 30, 40, and 50 kOe) as a function of $T$ (see fig. 4). $\Delta S$ is found to be positive for all the samples and it peaks around the respective MT region. This positive $\Delta S$ indicates inverse MCE, which has been observed in other Ni-Mn-Z samples~\cite{imce, krenke}. The peak value of the entropy change ($\Delta S_{max}$) is found to be highest (13.6 J /kg K) for the $x$ = 0.28 sample and it occurs around 249 K. For $x$ = 0.32 sample, the peak occurs even at higher $T$ (close to room temperature (274 K)) and with $\Delta S_{max}$ being 12.8 J/kg K. The $x$ = 0.24 sample shows $\Delta S_{max}$ of 6.8 J/kg K at around 219 K. The magnitudes of the MCE in all the samples are reasonably large, and they occur at temperatures which can be quite useful for magnetic refrigeration. In addition, one can actually {\it tune} the position of the MT (and hence the position of $\Delta S_{max}$) by simply varying the Ga concentration and without disturbing the concentration of the {\it magnetic element (here Mn)}.

\par
The peak in $\Delta S$ versus $T$ data is much sharper in case of $x$ = 0.32 and 0.28 samples as compared to 0.24 sample. The full-width at half maximum (FWHM) of the peaks are respectively 12.5, 6.4 and 6.9 K for $x$ = 0.24, 0.28 and 0.32 samples. The refrigeration capacity (RC = the area of the $\Delta S$ versus $T$ data calculated by FWHM $\times$ peak height) of a sample is a very important parameter for the development of MCE based refrigerator. For $x$ = 0.32 sample, RC is found to be 108 J/kg, which is a reasonably large quantity as far as the magneto-caloric materials are concerned. For $x$ = 0.24 and 0.28, RC is found to be 88.6 and 103.6 J/kg respectively.

\section{Discussion}

For pure Ni-Mn-Ga alloys, the MCE near the structural transition is often negative. Although, low field positive MCE has also been reported in certain Ni-Mn-Ga alloys. The positive MCE in Ni-Mn-Ga alloys is generally attributed to the magnetocrystalline anisotropy (through $H$ induced variant reorientation mechanism) or to the enhancement of $T_{MS}$ by $H$(see ~\cite{planes} and references therein). On the contrary, in Ni-Mn-Z (with Z= Sn, Sb, In), the MCE near $T_{MS}$ is always positive and can have very large magnitude. Due to the first order nature of MT, the transition region is highly metastable and both martensite and austenite can coexist. The positive MCE in these alloys is related to the martensite to austenite field-induced reverse phase transition. For the present samples, which are the solid solutions of Ni-Mn-Sn and Ni-Mn-Ga alloys, the MCE is found to be large and positive in all the compositions. This clearly indicates that the reverse transition is playing the key role toward the observed MCE. The mixed samples predominantly reflect the properties of Sn samples, even for 50\% Ga substitution ($x$ = 0.32). The signature of field induced transition is also clear from the signature of metamagnetism in the $M-H$ isotherms (fig. 3 (b)). 

\begin{figure}%
\centering
\includegraphics[width=6.5 cm]{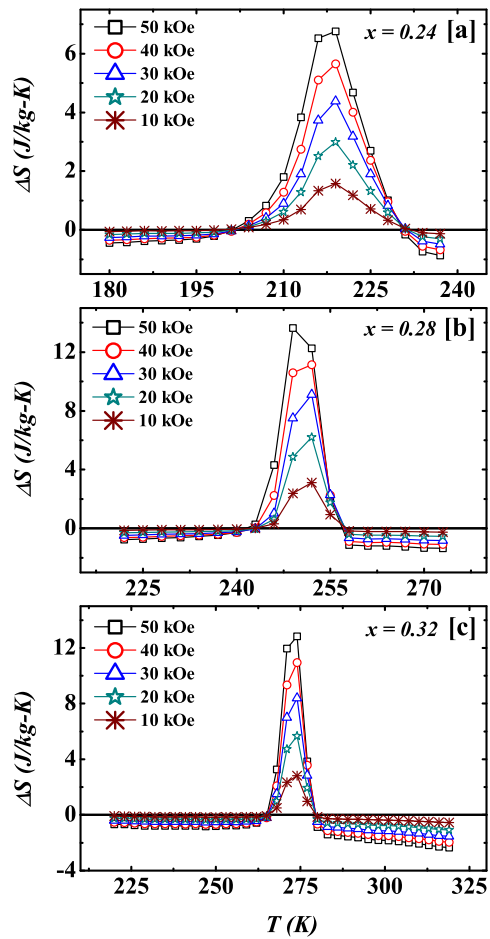}%
\caption{Change in entropy due to the application of different magnetic fields for $x$ = 0.24, 0.28 and 0.32 samples depicted in (a), (b) and (c) respectively.}%
\end{figure}

\par
The studied Ni-Mn-Sn-Ga samples resembles the Ni-Mn-Sn samples in some other magnetic properties. In this regard, the step like anomaly in the ZFCH $M(T)$ curve is worth noting. Such step like anomaly (marked by $T_B$ in fig. (2)) is connected to the development of some glassy magnetic state and it is also the blocking temperature for the observed exchange bias (EB) at low temperature ~\cite{eb, sc2}. The observed glassyness and EB are related to the interfacial coupling between the majority FM fraction and the incipient antiferromagnetic (AFM) fraction originating from the excess Mn atoms in the sample over the Heusler stoichiometry.

\par
The reduction in moment in the off-stoichiometry samples is perhaps related to the enhanced antiferromagnetic (AFM) correlation in the alloys. The excess Mn atoms in the alloys over the Heusler stoichiometry (X$_2$YZ) at the expense of Z atoms (for the present samples, Ni$_2$Mn$_{1.36}$Sn$_{0.64-x}$Ga$_{x}$ we have 36\% more Mn) ocuupy 4$c$ Wykoff position of the Heusler structure, while the orginal Mn atoms stay at 4$b$ Wykoff position~\cite{mc}. The 4$b$ and 4$c$ Mn atoms in a unit cell interact antiferromagnetically and reduces the net moment. Gallium doping makes the distance between Mn atoms at 4$b$ and 4$c$ sites shorter, thereby enhances this intersite AFM correlations, and consequently the net moment decreases. It is worth noting that the step like anomaly $T_B$ increases with Ga concentration. This is also an indication of enhanced AFM correlation, which is responsible for the EB and glassyness of the alloys. 

\begin{table}
\caption{Variation of cubic lattice parameter ($a_{cubic}$), bond distance between Mn atoms at 4$b$ and 4$c$ sites, valence electron concentration ($e/a$) saturation moment ($M_{sat}$) and martensitic transition temperature ($T_{MS}$) in Ni$_2$Mn$_{1.36}$Sn$_{0.64-x}$Ga$_{x}$ with $x$.} 

\vskip 0.2 cm
\centering
\begin{tabular}{cccccc}
\hline
\hline
$x$ & $a_{cubic}$(\AA) & [Mn(4$b$)-Mn(4$c$)] (\AA)& $e/a$ ratio & $M_{sat}$ ($\mu_B$/f.u.) & $T_{MS}$ (K)\\ 
\hline
0.00 & 5.99 & 2.995 & 8.02 & 2.74 & 110 \\
0.24 & 5.94 & 2.970 & 7.96 & 2.28 & 208 \\ 
0.28 & 5.93 & 2.965 & 7.95 & 2.24 & 231 \\ 
0.32 & 5.92 & 2.960 & 7.94 & 2.12 & 254 \\
\hline
\hline
\end{tabular}
\end{table}

\par
It is interesting to note that $T_{MS}$ of the samples increases with Ga doping, while $M_{sat}$ decreases (see table 1). Doping with Ga actually reduces the valence electron concentration ($e/a$), and on the basis of Hume Rothery mechanism one would expect a gradual decrease of $T_{MS}$ with $x$~\cite{vc1}. This anomalous behaviour of $T_{MS}$ possibly originates from the chemical pressure effect due to Ga doping, which can dominate over the effect related to $e/a$. It has been observed in many Ni-Mn-Z based alloys that hydrostatic pressure ($P$) can enhance the martensitic transition temperature~\cite{vc2, lm} by virtue of Clausius-Clapeyron relation, $dT/dP = T\Delta V/L$, where $\Delta V$ is the change in volume at the MT, and $L$ is the latent heat of transition. Since in MT, $\Delta V$ is negative, one would expect an enhanced $T_{MS}$ with $P$.

\par
In conclusion, we report the formation of a new series of ferromagnetic shape memory alloys by partial replacement of Sn by Ga in Ni$_2$Mn$_{1.36}$Sn$_{0.64}$. The Ga doping enhances $T_{MS}$ of the alloys considerably, and for 50\% Ga replacement, it reaches close to room temperature. The Ga doped samples retains most of the magnetic properties including ferromagnetism and the field-indeuced transition of the pure Sn sample. We observe very large value of MCE in the Ga doped sample, which occurs close to room temperature. The samples show large refrigeration capacity. The present samples can be a useful addition to the list of ferromagnetic shape memory alloys showing giant MCE near room temperature. 

\par
The present work is financially supported by CSIR, India.


\end{document}